\documentclass[submission,copyright,creativecommons]{eptcs}
\providecommand{\event}{MARS 2015} 

\usepackage{isabelle,isabellesym}
\usepackage[pdftex]{graphicx}

\usepackage{microtype}
\usepackage{latexsym}
\usepackage{proof}
\usepackage{mathpartir}
\usepackage{relsize}
\usepackage{tabto}
\usepackage{caption}
\usepackage{subcaption}
\usepackage{color}
\definecolor{lcol}{rgb}{0,0,0.5}

\usepackage{pdfsetup}

\isabellestyle{it}

\newif \ifDraft         \Draftfalse

\Draftfalse

\ifDraft
  \newcommand{\Comment}[1]{\textbf{\textsl{#1}}}
  \newcommand{\FIXME}[1]{\textbf{\textsl{\colorbox{yellow}{FIXME:} #1}}}
  \newcommand{\TODO}[1]{\textbf{\textsl{TODO: #1}}}
\else
  \newcommand{\Comment}[1]{\relax}
  \newcommand{\FIXME}[1]{\relax}
  \newcommand{\TODO}[1]{\relax}
\fi

\newcommand{\toby}[1]{\Comment{#1 [Toby]}}
\newcommand{\sid}[1]{\Comment{#1 [Sidney]}}
\newcommand{\fsop}[1]{\texttt{#1()}}

\providecommand{\urlalt}[2]{\href{#1}{#2}}
\providecommand{\doi}[1]{doi:\urlalt{http://dx.doi.org/#1}{#1}}

\newcommand{\ptitle}{Specifying a Realistic File System}

\title{\ptitle}
\author{Sidney Amani \qquad Toby Murray
\institute{NICTA and University of New South Wales, Australia}
}
\def\titlerunning{\ptitle}
\def\authorrunning{Sidney Amani \& Toby Murray}

\usepackage{xspace}

\begin{document}
\maketitle

\begin{abstract}

  We present the most interesting elements of the correctness
  specification of BilbyFs, a performant Linux flash
  file system.
  The BilbyFs specification supports asynchronous writes, a feature that has
  been overlooked by several file system verification projects, and has been
  used to verify the correctness of BilbyFs's \fsop{fsync} C implementation.
  It makes use of nondeterminism to be concise and is shallowly-embedded
  in higher-order logic.

\end{abstract}

\section{Introduction\label{s:intro}}

File systems have been a target of software verification
research for decades~\cite{Morgan_Sufrin_84,Bevier_CT_95,Joshi_Holzmann_07}.
While recent research has increased the fidelity of the specifications
against which file system implementations are to be verified,
the gap remains large between the details and features covered
by existing verification work, on the one hand, and those for production
file systems as found in e.g. the Linux kernel, on the other.

A common simplification has been to omit \emph{asynchronous writes}
from the file system specification and implementation~\cite{Ernst_SHPR_13,Hesselink_Lali_09,Chen_ZCCKZ_15}.
File systems typically buffer state-updates
in-memory, so that updates to the physical storage medium can be batched
to improve throughput. Updates are propagated to the storage medium
periodically, or explicitly via the file system \fsop{fsync} operation, meaning
that they occur asynchronously.
This asynchrony is crucial for performance but complicates
the file system specification and implementation.

In this paper, we describe the most interesting elements of
the correctness specification of BilbyFs.
BilbyFs is a custom flash file system we developed that runs as a Linux kernel module
and supports asynchronous writes, and whose
\fsop{fsync} implementation has been verified as functionally correct against the correctness specification we present here.
To make the verification tractable, BilbyFs enforces sequential
execution of the file system code by acquiring a global lock
before invoking an operation.
While out of scope for this paper, BilbyFs performs
comparably to mainstream flash file systems like
UBIFS~\cite{Hunter_08} and JFFS2~\cite{Woodhouse_03}, and its design strikes a balance between the simplicity
of JFFS2 and the runtime performance and reliability of UBIFS.

The BilbyFs specification is complementary to that of Chen et al.~\cite{Chen_ZCCKZ_15}
who recently verified FSCQ, a crash-safe user-space file system 
implemented in Haskell. FSCQ performs asynchronous writes only \emph{within} each individual
file system operation, and synchronously waits for writes to complete at the end of
each operation. In contrast, BilbyFs allows entire sequences of file system operations to occur
asynchronously. Their specification describes
FSCQ's FUSE interface, which is closely aligned with the POSIX interface expected by application 
programs, and is specified as a series of Hoare triples over the top-level functions. 
The BilbyFs specification is a functional program in Isabelle/HOL (\S\ref{s:formalisation})
that describes the interface BilbyFs provides to the Linux kernel's Virtual File
system Switch (VFS), which is at a different level of abstraction than FUSE.
To support fully asynchronous operations, in contrast to Chen et al.'s, the BilbyFs specification
explicitly
separates the in-memory and on-medium file system state (\S\ref{s:absspec})
which allows us to specify the effect of the \fsop{fsync} operation (\S\ref{s:ops})
and reduce the gap between realistic file systems and verified ones (\S\ref{s:limitations}).

\begin{figure}
\centering
\begin{minipage}{.5\textwidth}
  \centering
  \includegraphics[width=.7\linewidth]{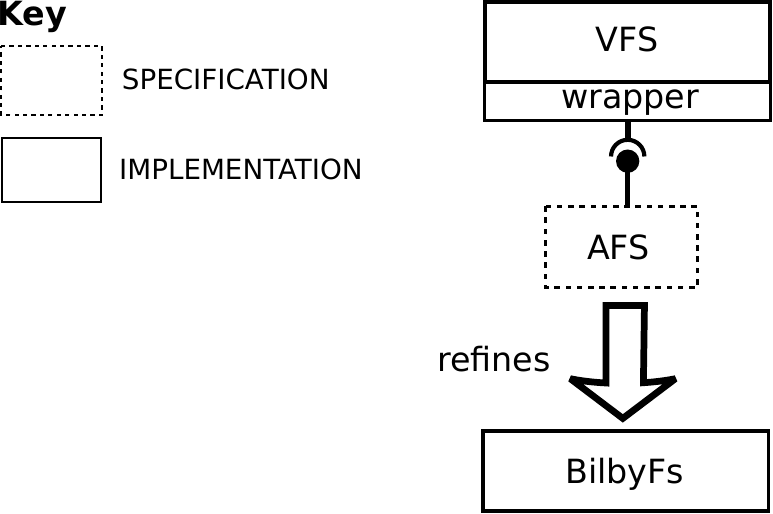}
  \captionof{figure}{Correctness specification overview}
  \label{f:afsoverview}
\end{minipage}%
\begin{minipage}{.5\textwidth}
  \centering
  \vspace{0.1cm}
  \includegraphics[width=1\linewidth]{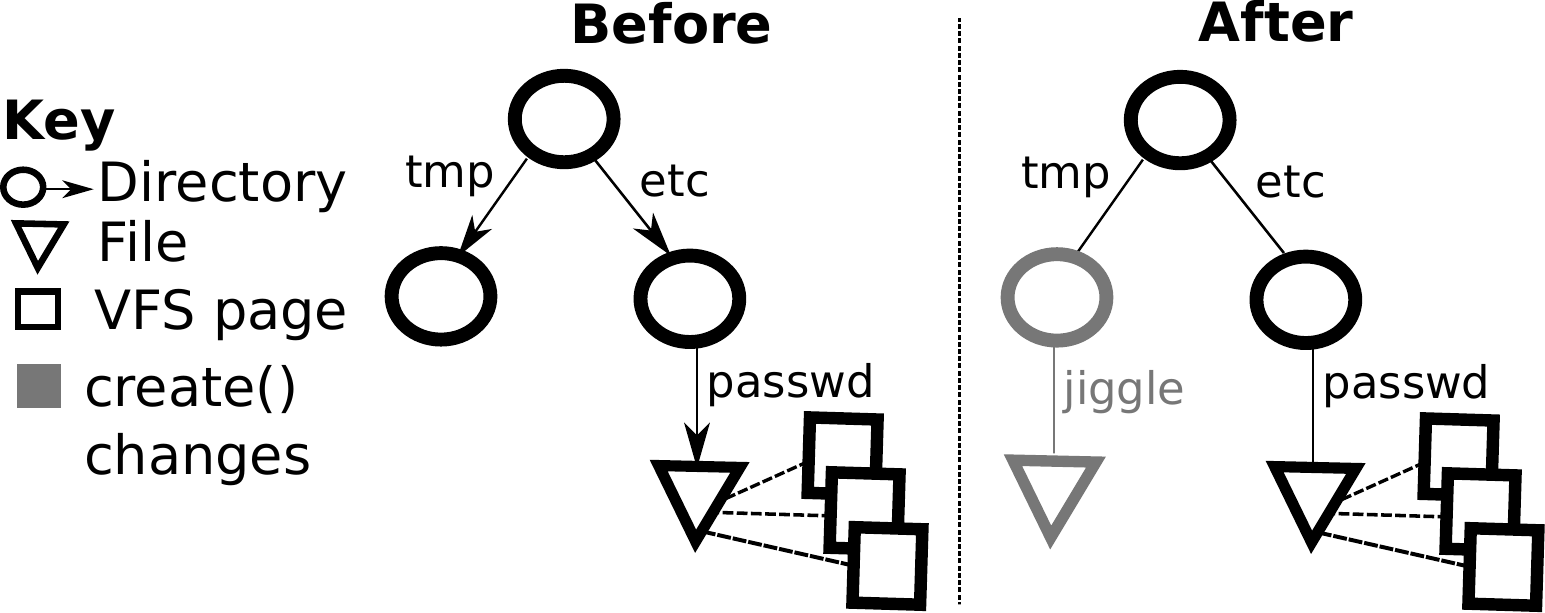}
  \vspace{0.1cm}
  \captionof{figure}{The effect of the \fsop{create} operation}
  \label{f:abscreate}
\end{minipage}
\end{figure}

\section{Formalisation\label{s:formalisation}}
\begin{isabellebody}%
\setisabellecontext{CorrectnessSpec}%
\isadelimtheory
\endisadelimtheory
\isatagtheory
\endisatagtheory
{\isafoldtheory}%
\isadelimtheory
\endisadelimtheory
\isadelimproof
\endisadelimproof
\isatagproof
\endisatagproof
{\isafoldproof}%
\isadelimproof
\endisadelimproof
\begin{isamarkuptext}%
\FIXME{Explain why we chose this formalism: no reasonining about concurrency, we need
        shallow embedding to ease the proofs, having a monad allows syntactic matching
       for our refinement framework}
Our objective is to prove the \emph{functional correctness} of BilbyFs,
specifically that every behaviour exhibited by the implementation is captured by
its specification (i.e. formal refinement~\cite{deRoever_Engelhardt:DR}).
The BilbyFs implementation uses a global lock to avoid all concurrency,
meaning that its specification can be entirely sequential.
This specification should be concise enough to be easily
audited to ensure it captures the intended behaviour of BilbyFs.

We chose to shallowly-embed the correctness specification of BilbyFs in
Isabelle/HOL~\cite{Nipkow_PW:Isabelle}.
Our specification is formalised in the \emph{nondeterminism monad}, inspired by
the nondeterministic state monad 
of Cock et al.\cite{Cock_KS_08}, where computations are modelled as sets of values, one for each
possible execution and associated return-value of the computation. A computation
that returns results of type \isa{{\isasymsigma}} is modelled as a value of type \mbox{\isa{{\isasymsigma}\ set}}:
singleton sets model deterministic computations; the empty set models
computations whose semantics are not well-defined.

The primitive computation \isa{return\ x} simply yields the result
\isa{x}: \isa{return\ x\ {\isasymequiv}\ {\isacharbraceleft}x{\isacharbraceright}}.
The ``bind'' operation, written \isa{{\isachargreater}{\isachargreater}{\isacharequal}}, sequences two
computations together. \mbox{\isa{f\ {\isachargreater}{\isachargreater}{\isacharequal}\ g}} is the specification
that for each result~\isa{r} of the computation~\isa{f}, executes~\isa{g} passing
\isa{r} as \isa{g}'s argument: \isa{f\ {\isachargreater}{\isachargreater}{\isacharequal}\ g\ {\isasymequiv}\ {\isasymUnion}\isactrlbsub r{\isasymin}f\isactrlesub \ g\ r}.
The nondeterministic choice~\isa{x\ {\isasymsqinter}\ y} between two computations~\isa{x} and \isa{y}
is defined trivially: \isa{x\ {\isasymsqinter}\ y\ {\isasymequiv}\ x\ {\isasymunion}\ y}.
We write \mbox{\isa{do\ x\ {\isasymleftarrow}\ f{\isacharsemicolon}\ g\ x\ od}} as sugar for \isa{f\ {\isachargreater}{\isachargreater}{\isacharequal}\ g}, in
the style of Haskell.

The do-notation makes the specification readable by ordinary programmers. Being
shallowly-embedded in Isabelle/HOL, the specification is more straightforward
to reason about than if it were deeply-embedded. Finally, following~\cite{Cock_KS_08},
this formalism supports a scalable, compositional framework for proving refinement.

Unlike Cock et al., we eschew a state monad for the AFS. Our specifications
are simple enough that passing the file system state explicitly adds
little overhead. Further, we found that a state monad makes our specifications
harder to read while imposing extra mental overhead due to having to unpack and repack
the state when calling into sub-modules that only touch a small part of it.

\section{File System Abstraction\label{s:absspec}}

Having described the formalism of our specification, we now
describe more precisely the level of abstraction at which we
specify file system behaviours and we present our abstract
model of the file system state used to specify asynchronous
writes.

Modern operating systems include a variety of file systems.
Each in-kernel file system does not directly provide functionality to the user,
but instead interacts via the kernel's VFS that provides a
common interface to all file systems. Thus the top-level operations provided by
BilbyFs, and described by its correctness specification,
are those expected by the VFS. These
include a total of 16 file system operations.

\fsop{create}, \fsop{unlink}, \fsop{mkdir} and \fsop{rmdir} for respectively creating and removing both
files and directories, as well as \fsop{readpage}, \fsop{write\_begin}, \fsop{write\_end}, \fsop{readdir} for reading and writing files and listing directories.
\fsop{lookup} finds an inode number by name and reads it from disk.
\fsop{rename} renames or move a file or directory.
\fsop{symlink} creates a symbolic link and \fsop{follow\_link} reads the path stored in a symbolic link.
\fsop{link} creates a hardlink to a file.
\fsop{setattr} and \fsop{getattr} for accessing and altering attributes of file and directories.
Finally, \fsop{evict\_inode} is called when an file is evicted from the cache and
\fsop{fsync} synchronises in-memory updates to storage.

\autoref{f:afsoverview}
shows how the correctness specification, which we call the 
\emph{Abstract File system Specification} (AFS), relates to the
Linux VFS and to the BilbyFs implementation.
A small C wrapper that sits between the Linux VFS and BilbyFs
decouples VFS interfaces from the Linux kernel caches and serialises
the execution of file system operations (see \autoref{s:limitations}).
BilbyFs operations are invoked by the VFS through the wrapper;
the AFS describes concisely how these operations should behave.

The BilbyFs functional correctness proof, completed for the \fsop{fsync}
operation, establishes
that the BilbyFs implementation correctly implements the AFS, by proving that the
former formally \emph{refines}~\cite{deRoever_Engelhardt:DR} the latter. 

\paragraph{Modelling the File System State}
Internally, a file system implements its hierarchical structure on disk in linked
objects called \emph{inodes}, which are file system specific. 
Similarly to previously-mentioned work~\cite{Ernst_SHPR_13,Hesselink_Lali_09},
the AFS models the file system state as the type \isa{afs{\isacharunderscore}map}, which is a
synonym for a (partial) mapping from \emph{inode numbers} (32-bit integers that
uniquely identify inodes) to \isa{afs{\isacharunderscore}inode} objects.
\isa{afs{\isacharunderscore}inode}s abstractly model the
on-medium representation of BilbyFs inodes.%
\end{isamarkuptext}%
\isamarkuptrue%
\isacommand{type{\isacharunderscore}synonym}\isamarkupfalse%
\ afs{\isacharunderscore}map\ {\isacharequal}\ {\isachardoublequoteopen}ino\ {\isasymrightharpoonup}\ afs{\isacharunderscore}inode{\isachardoublequoteclose}\isanewline
\isanewline
\isacommand{datatype}\isamarkupfalse%
\ afs{\isacharunderscore}inode{\isacharunderscore}type\ {\isacharequal}\isanewline
\ \ IDir\ {\isachardoublequoteopen}filename\ {\isasymrightharpoonup}\ ino{\isachardoublequoteclose}\isanewline
{\isacharbar}\ IFile\ {\isachardoublequoteopen}vfspage\ list{\isachardoublequoteclose}\isanewline
\isanewline
\isacommand{record}\isamarkupfalse%
\ afs{\isacharunderscore}inode\ {\isacharequal}\isanewline
\ \ i{\isacharunderscore}type\ {\isacharcolon}{\isacharcolon}\ {\isachardoublequoteopen}afs{\isacharunderscore}inode{\isacharunderscore}type{\isachardoublequoteclose}\isanewline
\ \ i{\isacharunderscore}ino\ {\isacharcolon}{\isacharcolon}\ {\isachardoublequoteopen}ino{\isachardoublequoteclose}\isanewline
\ \ i{\isacharunderscore}nlink\ {\isacharcolon}{\isacharcolon}\ {\isachardoublequoteopen}{\isadigit{3}}{\isadigit{2}}\ word{\isachardoublequoteclose}\isanewline
\ \ i{\isacharunderscore}size\ {\isacharcolon}{\isacharcolon}\ {\isachardoublequoteopen}{\isadigit{6}}{\isadigit{4}}\ word{\isachardoublequoteclose}%
\begin{isamarkuptext}%
\vspace{-3ex}\ \ $\ldots$%
\end{isamarkuptext}%
\isamarkuptrue%
\begin{isamarkuptext}%
An \isa{afs{\isacharunderscore}inode} represents either a directory or a file, and the \isa{i{\isacharunderscore}type}
 field stores both what kind of object it represents as well as the content it holds
 in each case. Each directory inode holds a mapping from filenames (byte strings) to
 associated inode numbers. Each file inode links to the data pages that hold its file's contents.

 This abstraction allows the AFS to be as simple as possible while comprehensively
 capturing the behaviour of the various file system operations.
 In practice, inode implementations are a source of complexity and defects~\cite{Lu_AAL_14}
 for file systems. For instance the inode contents described above, which the 
 AFS models as stored directly in the \isa{i{\isacharunderscore}type} field of the \isa{afs{\isacharunderscore}inode} object, 
 will in practice be stored in separate data structures on the medium to which the inode links,
 often via multiple levels of indirection.

 The AFS represents a file's contents as a list of VFS pages (each of which is
 a list of bytes). An alternative would have been to use just a single flat list of
 bytes. However, we chose the former because the VFS layer interacts closely with the
 Linux page cache (i.e. memory management subsystem) which is
 directly reflected in the VFS API that BilbyFs implements.

 Directory inodes are usually implemented as a collection of directory
 entries stored in a file. The AFS instead models directory contents as 
 a mapping from filenames
 to inode identifiers: the inode identifier for a particular filename identifies
 the file inode that implements the file with that name.
 In file systems like BilbyFs that support \emph{hardlinks},
 the same file inode may be linked to by multiple filenames (in possibly different
 directories). Like most file systems, BilbyFs restricts hardlinks to point only to
 file inodes, to ensure that the file system's logical structure is acyclic.
 
\paragraph{AFS Invariant\label{s:invariant}}

The aforementioned requirement on hardlinks is encoded as part of the global invariant
of the AFS, which we describe only briefly because (like the AFS state itself) it is
very similar to that of~\cite{Ernst_SHPR_12}.
The invariant assertion includes the following requirements of the \isa{afs{\isacharunderscore}map} structure.
(1) A root inode exists; (2) for each directory inode, there is only one directory with
an entry pointing to that inode --- i.e. no hardlinks on directories;
(3) the \isa{i{\isacharunderscore}ino} field of each inode matches the \isa{afs{\isacharunderscore}map} mapping;
(4) the \isa{i{\isacharunderscore}nlink} field of each inode correctly tracks the number of links to
that inode; and (5) the \isa{i{\isacharunderscore}size} field of each file inode is consistent with
the number of VFS pages attached to the inode.

 \paragraph{Modelling File System Updates\label{s:fsops}}
 
 The AFS's abstract representation of the file system state makes
 describing the effects of the file system operations easy,
 and in turn facilitates auditing of the specification.
 
 \autoref{f:abscreate} depicts the changes to the \isa{afs{\isacharunderscore}map}
 structure when the \fsop{create} operation is called to create a new
 file \texttt{jiggle} in the \texttt{/tmp} directory. Grey nodes are those
 altered or added by the operation.  Creating a file involves adding
 a file inode and a link to it in the parent directory.
 Directories are pictured as circles with arrows denoting each entry in the
 directory. Small triangles denote files and are linked to VFS pages shown
 as tiny squares.
 The newly created file contains no data so no VFS page is attached to its inode.

 The effect on the \isa{afs{\isacharunderscore}map} structure~\isa{m} of creating a new file \isa{{\isacharprime}{\isacharprime}jiggle{\isacharprime}{\isacharprime}} in a directory (whose inode is)
 \isa{itmp}, by installing the new file inode \isa{ijiggle}, is specified by
 the following single line of Isabelle/HOL\footnote{Note that the specification for
 the \fsop{create} operation (see \autoref{fig:create-corrspec}) is a lot
 more complex than this single line, because it needs to also incorporate error checking
and handling, and interaction with the Linux VFS.
}.%
\end{isamarkuptext}%
\isamarkuptrue%
\begin{isamarkuptext}%
\isa{m{\isasymlangle}i{\isacharunderscore}ino\ itmp\ {\isasymmapsto}\ i{\isacharunderscore}dir{\isacharunderscore}upd\ {\isacharparenleft}{\isasymlambda}dir{\isachardot}\ dir{\isasymlangle}{\isacharprime}{\isacharprime}jiggle{\isacharprime}{\isacharprime}\ {\isasymmapsto}\ i{\isacharunderscore}ino\ ijiggle{\isasymrangle}{\isacharparenright}\ itmp{\isacharcomma}\ i{\isacharunderscore}ino\ ijiggle\ {\isasymmapsto}\ ijiggle{\isasymrangle}}

We write \isa{m{\isasymlangle}a\ {\isasymmapsto}\ b{\isasymrangle}} to denote updating the (partial) map
\isa{m} such that \isa{a} maps to \isa{b}.

Each transformation to the file system state may be captured by a function
of type~\mbox{\isa{afs{\isacharunderscore}map\ {\isasymRightarrow}\ afs{\isacharunderscore}map}}. We call such functions \emph{file system updates}. 
We exploit this idea in the following section,
where we describe the model of asynchronous writes used in the AFS.%
\end{isamarkuptext}%
\isamarkuptrue%
\isadelimtheory
\endisadelimtheory
\isatagtheory
\endisatagtheory
{\isafoldtheory}%
\isadelimtheory
\endisadelimtheory
\end{isabellebody}%

\paragraph{Specifying Asynchronous Writes\label{s:asyncwrites}}

\begin{isabellebody}%
\setisabellecontext{AsyncWriteSpec}%
\isadelimtheory
\endisadelimtheory
\isatagtheory
\endisatagtheory
{\isafoldtheory}%
\isadelimtheory
\endisadelimtheory
\isadelimproof
\endisadelimproof
\isatagproof
\endisatagproof
{\isafoldproof}%
\isadelimproof
\endisadelimproof
\begin{isamarkuptext}%
The \isa{afs{\isacharunderscore}map} type models the state of the file system as stored
on the physical storage medium, which in the case of BilbyFs is the raw flash
storage device, and updates to the storage medium are simply transformations:
\isa{afs{\isacharunderscore}map\ {\isasymRightarrow}\ afs{\isacharunderscore}map}.  Like many other realistic file systems,
although the effects of file system operations like \fsop{create},
\fsop{unlink} etc., become visible as soon as those invocations return,
the actual storage medium update that they implement may not be applied
until some point in the future, for instance when \fsop{fsync}
is next invoked. Thus writes to the storage medium are performed
\emph{asynchronously}, an essential feature of file systems since
the original UNIX~\cite{Thompson_78}.
Storage medium updates are therefore buffered in-memory, allowing
operations like \fsop{create} to return straightaway, without incurring
the cost writing to the storage medium. For the file system correctness
specification, this implies
that the in-memory file system state and the state of the physical storage
medium (\isa{afs{\isacharunderscore}map} in the AFS) need to be distinguished, especially if
the semantics of operations like \fsop{fsync} are to be precisely specified. 

As mentioned in \autoref{s:intro}, several file system
models~\cite{Ernst_SHPR_13,Hesselink_Lali_09,Chen_ZCCKZ_15} overlook this
requirement.
In the AFS for BilbyFs, the pending writes buffered in-memory are modelled
as a sequence of file system transformations, each of type \isa{afs{\isacharunderscore}map\ {\isasymRightarrow}\ afs{\isacharunderscore}map}. 
The global state of the specification is captured by the type \isa{afs{\isacharunderscore}state}.
Besides the state of the physical medium, of type \isa{afs{\isacharunderscore}map}; and in-memory pending updates,
of type \isa{{\isacharparenleft}afs{\isacharunderscore}map\ {\isasymRightarrow}\ afs{\isacharunderscore}map{\isacharparenright}\ list}; \isa{afs{\isacharunderscore}state} also includes a boolean flag that tracks whether
the file system has been placed into \emph{read-only mode}, which can occur for instance
if a disk failure arises; as well as a record of the current time, which is used for instance
when updating the timestamps on an inode that track e.g. the last time it was modified.%
\end{isamarkuptext}%
\isamarkuptrue%
\isacommand{record}\isamarkupfalse%
\ afs{\isacharunderscore}state\ {\isacharequal}\isanewline
\ \ a{\isacharunderscore}is{\isacharunderscore}readonly\ {\isacharcolon}{\isacharcolon}\ {\isachardoublequoteopen}bool{\isachardoublequoteclose}\isanewline
\ \ a{\isacharunderscore}current{\isacharunderscore}time\ {\isacharcolon}{\isacharcolon}\ {\isachardoublequoteopen}time{\isachardoublequoteclose}\isanewline
\ \ a{\isacharunderscore}medium{\isacharunderscore}afs\ {\isacharcolon}{\isacharcolon}\ {\isachardoublequoteopen}afs{\isacharunderscore}map{\isachardoublequoteclose}\isanewline
\ \ a{\isacharunderscore}medium{\isacharunderscore}updates\ {\isacharcolon}{\isacharcolon}\ {\isachardoublequoteopen}{\isacharparenleft}afs{\isacharunderscore}map\ {\isasymRightarrow}\ afs{\isacharunderscore}map{\isacharparenright}\ list{\isachardoublequoteclose}%
\begin{isamarkuptext}%
To model the idea that file system modifications become visible straightaway,
even when they have not yet been applied to the physical storage medium, the AFS needs 
a way to calculate the (hypothetical) file system state that includes both the physical medium
and the pending in-memory updates, i.e. the state that \emph{would} arise
if all of those updates were applied to the medium. It is this hypothetical state that
must be considered, for instance, when the \fsop{unlink} operation is invoked to remove
an inode that was previously \fsop{create}ed but hasn't yet been \fsop{fsync}ed to disk.
It is calculated by the function \isa{updated{\isacharunderscore}afs}, that makes use of the standard
\isa{fold} function to apply the in-memory updates to the medium state:

\begin{isabelle}%
updated{\isacharunderscore}afs\ afs{\isacharunderscore}state\ {\isasymequiv}\isanewline
fold\ {\isacharparenleft}{\isasymlambda}x{\isachardot}\ x{\isacharparenright}\ {\isacharparenleft}a{\isacharunderscore}medium{\isacharunderscore}updates\ afs{\isacharunderscore}state{\isacharparenright}\ {\isacharparenleft}a{\isacharunderscore}medium{\isacharunderscore}afs\ afs{\isacharunderscore}state{\isacharparenright}%
\end{isabelle}

An operation like \fsop{create} that updates the file system state may buffer the updates
it performs in memory, or (if the in-memory buffer is full) it may cause preceding updates to be 
applied to the storage medium. Given that the size of the in-memory buffer is below the
level of abstraction of the AFS, the precise number of updates that may be propagated to 
the storage medium could vary upwards from zero. The AFS captures this effect via the 
following helper function, which nondeterministically splits the list of updates into
two parts: \isa{to{\isacharunderscore}apply}, the updates to be applied to the medium; and \isa{rem},
the remainder. It then applies the updates in \isa{to{\isacharunderscore}apply} and updates the in-memory
list of pending updates to \isa{rem}.

\begin{isabelle}%
afs{\isacharunderscore}apply{\isacharunderscore}updates{\isacharunderscore}nondet\ afs\ {\isasymequiv}\isanewline
do\ {\isacharparenleft}to{\isacharunderscore}apply{\isacharcomma}\ rem{\isacharparenright}\ {\isasymleftarrow}\ {\isacharbraceleft}{\isacharparenleft}t{\isacharcomma}\ r{\isacharparenright}\ {\isacharbar}\ t\ {\isacharat}\ r\ {\isacharequal}\ a{\isacharunderscore}medium{\isacharunderscore}updates\ afs{\isacharbraceright}{\isacharsemicolon}\isanewline
\ \ \ return\isanewline
\isaindent{\ \ \ \ }{\isacharparenleft}afs{\isasymlparr}a{\isacharunderscore}medium{\isacharunderscore}afs\ {\isacharcolon}{\isacharequal}\ fold\ {\isacharparenleft}{\isasymlambda}x{\isachardot}\ x{\isacharparenright}\ to{\isacharunderscore}apply\ {\isacharparenleft}a{\isacharunderscore}medium{\isacharunderscore}afs\ afs{\isacharparenright}{\isacharcomma}\isanewline
\isaindent{\ \ \ \ {\isacharparenleft}afs\ \ \ }a{\isacharunderscore}medium{\isacharunderscore}updates\ {\isacharcolon}{\isacharequal}\ rem{\isasymrparr}{\isacharparenright}\isanewline
od%
\end{isabelle}

The following helper function \isa{afs{\isacharunderscore}update} then generically specifies the process
for updating the file system state, and is used in the specifications of the various
file system operations (see e.g. \fsop{create} in \autoref{fig:create-corrspec}). 
It takes an update function~\isa{upd}
of type \isa{afs{\isacharunderscore}map\ {\isasymRightarrow}\ afs{\isacharunderscore}map}. It adds it to the back of the list of in-memory updates
and then calls \isa{afs{\isacharunderscore}apply{\isacharunderscore}updates{\isacharunderscore}nondet}. If after \isa{afs{\isacharunderscore}apply{\isacharunderscore}updates{\isacharunderscore}nondet}
returns, the list of in-memory updates is empty, then \isa{afs{\isacharunderscore}apply{\isacharunderscore}updates{\isacharunderscore}nondet}
caused all in-memory updates (including \isa{upd}) to be propagated to the storage medium,
in which case the modification must report \isa{Success}. Otherwise, it might succeed
(if, for instance, the new update is simply buffered in-memory without touching the storage
medium), or report an appropriate error (because a write to the medium failed, or a memory allocation
error occurred etc.). If an error is reported, the new update~\isa{upd} is forgotten,
ensuring operations that report an error do not modify the (combined in-memory and on-medium)
file system state.

\begin{isabelle}%
afs{\isacharunderscore}update\ afs\ upd\ {\isasymequiv}\isanewline
do\ afs\ {\isasymleftarrow}\isanewline
\isaindent{do\ }afs{\isacharunderscore}apply{\isacharunderscore}updates{\isacharunderscore}nondet\isanewline
\isaindent{do\ \ }{\isacharparenleft}afs{\isasymlparr}a{\isacharunderscore}medium{\isacharunderscore}updates\ {\isacharcolon}{\isacharequal}\ a{\isacharunderscore}medium{\isacharunderscore}updates\ afs\ {\isacharat}\ {\isacharbrackleft}upd{\isacharbrackright}{\isasymrparr}{\isacharparenright}{\isacharsemicolon}\isanewline
\ \ \ \textsf{if}\ a{\isacharunderscore}medium{\isacharunderscore}updates\ afs\ {\isacharequal}\ {\isacharbrackleft}{\isacharbrackright}\ \textsf{then}\ return\ {\isacharparenleft}afs{\isacharcomma}\ Success\ {\isacharparenleft}{\isacharparenright}{\isacharparenright}\isanewline
\isaindent{\ \ \ }\textsf{else}\ return\ {\isacharparenleft}afs{\isacharcomma}\ Success\ {\isacharparenleft}{\isacharparenright}{\isacharparenright}\ {\isasymsqinter}\isanewline
\isaindent{\ \ \ \textsf{else}\ }nondet{\isacharunderscore}error\ {\isacharbraceleft}eIO{\isacharcomma}\ eNoSpc{\isacharcomma}\ eNoMem{\isacharbraceright}\isanewline
\isaindent{\ \ \ \textsf{else}\ \ }{\isacharparenleft}{\isasymlambda}e{\isachardot}\ {\isacharparenleft}afs{\isasymlparr}a{\isacharunderscore}medium{\isacharunderscore}updates\ {\isacharcolon}{\isacharequal}\ butlast\ {\isacharparenleft}a{\isacharunderscore}medium{\isacharunderscore}updates\ afs{\isacharparenright}{\isasymrparr}{\isacharcomma}\isanewline
\isaindent{\ \ \ \textsf{else}\ \ {\isacharparenleft}{\isasymlambda}e{\isachardot}\ \ }Error\ e{\isacharparenright}{\isacharparenright}\isanewline
od%
\end{isabelle}

Importantly, this specification requires that no updates get lost when an error
occurs: each is either applied (in order), or is  still in the list of pending updates
(in order).  It also requires the BilbyFs implementation not to report an 
error if it succeeds in propagating all updates to disk. Thus the implementation
cannot attempt to allocate memory, for instance, after successfully writing to disk. In practice,
file system implementations structure each operation such that all resource allocation
(and other actions that could potentially fail) occur early, so no potentially-failing
operation needs to be performed after successfully updating the storage medium.

The \isa{afs{\isacharunderscore}update} definition is the heart of how the AFS specifies asynchronous
file system operations while keeping the AFS concise and readable. In the following
section, we present specifications of the most interesting top-level file system operations 
from the AFS.%
\end{isamarkuptext}%
\isamarkuptrue%
\isadelimtheory
\endisadelimtheory
\isatagtheory
\endisatagtheory
{\isafoldtheory}%
\isadelimtheory
\endisadelimtheory
\end{isabellebody}%

\section{Specifying File System Operations\label{s:ops}}
\paragraph{Specifying \fsop{create}}
\begin{isabellebody}%
\setisabellecontext{CreateSpec}%
\isadelimtheory
\endisadelimtheory
\isatagtheory
\endisatagtheory
{\isafoldtheory}%
\isadelimtheory
\endisadelimtheory
\isadelimproof
\endisadelimproof
\isatagproof
\endisatagproof
{\isafoldproof}%
\isadelimproof
\endisadelimproof
\begin{isamarkuptext}%
The specification for the \fsop{create} operation is shown in \autoref{fig:create-corrspec}.
Recall that BilbyFs's top-level operations, like \fsop{create}, 
are those expected by the Linux VFS. Since the VFS interacts with a range of
different file systems, each of which may have its own custom inode
format, the VFS provides a common inode abstraction, called a \emph{vnode}. Top-level
file system operations invoked by the VFS often take vnodes as their arguments and return
updated vnodes in their results. The \isa{afs{\isacharunderscore}inode} structure mentioned in
\autoref{s:absspec} is very similar to the generic vnode structure of the VFS.

Much of the complexity of \autoref{fig:create-corrspec} comes from error checking
and handling, as well as specifying the correct interaction with the VFS (e.g.
conversion from vnodes to \isa{afs{\isacharunderscore}inode}s).
\begin{figure}[tb]
\newcounter{afscreate}
\let\oldisanewline\isanewline
\renewcommand\isanewline{\oldisanewline\stepcounter{afscreate}\texttt{\arabic{afscreate}}\tabto{.75cm}}
\begin{isabelle}%
afs{\isacharunderscore}create\ afs\ vdir\ name\ mode\ vnode\ {\isasymequiv}\isanewline
\textsf{if}\ a{\isacharunderscore}is{\isacharunderscore}readonly\ afs\ \textsf{then}\ return\ {\isacharparenleft}{\isacharparenleft}afs{\isacharcomma}\ vdir{\isacharcomma}\ vnode{\isacharparenright}{\isacharcomma}\ Error\ eRoFs{\isacharparenright}\isanewline
\textsf{else}\ do\ r\ {\isasymleftarrow}\ afs{\isacharunderscore}init{\isacharunderscore}inode\ afs\ vdir\ vnode\ {\isacharparenleft}mode\ {\isacharbar}{\isacharbar}\ s{\isacharunderscore}IFREG{\isacharparenright}{\isacharsemicolon}\isanewline
\isaindent{\textsf{else}\ }\ \ \ \textsf{case}\ r\ \textsf{of}\ Error\ {\isacharparenleft}afs{\isacharcomma}\ vnode{\isacharparenright}\ {\isasymRightarrow}\ return\ {\isacharparenleft}{\isacharparenleft}afs{\isacharcomma}\ vdir{\isacharcomma}\ vnode{\isacharparenright}{\isacharcomma}\ Error\ eNFile{\isacharparenright}\isanewline
\isaindent{\textsf{else}\ \ \ \ }{\isacharbar}\ Success\ {\isacharparenleft}afs{\isacharcomma}\ vnode{\isacharparenright}\ {\isasymRightarrow}\isanewline
\isaindent{\textsf{else}\ \ \ \ {\isacharbar}\ \ \ }do\ r\ {\isasymleftarrow}\ read{\isacharunderscore}afs{\isacharunderscore}inode\ afs\ {\isacharparenleft}v{\isacharunderscore}ino\ vdir{\isacharparenright}{\isacharsemicolon}\isanewline
\isaindent{\textsf{else}\ \ \ \ {\isacharbar}\ \ \ }\ \ \ \textsf{case}\ r\ \textsf{of}\ Error\ e\ {\isasymRightarrow}\ return\ {\isacharparenleft}{\isacharparenleft}afs{\isacharcomma}\ vdir{\isacharcomma}\ vnode{\isacharparenright}{\isacharcomma}\ Error\ e{\isacharparenright}\isanewline
\isaindent{\textsf{else}\ \ \ \ {\isacharbar}\ \ \ \ \ \ }{\isacharbar}\ Success\ dir\ {\isasymRightarrow}\isanewline
\isaindent{\textsf{else}\ \ \ \ {\isacharbar}\ \ \ \ \ \ {\isacharbar}\ \ \ }do\ r\ {\isasymleftarrow}\ return\ {\isacharparenleft}Success\ {\isacharparenleft}i{\isacharunderscore}dir{\isacharunderscore}update\ {\isacharparenleft}{\isasymlambda}d{\isachardot}\ d{\isasymlangle}{\isasymalpha}wa\ name\ {\isasymmapsto}\ v{\isacharunderscore}ino\ vnode{\isasymrangle}{\isacharparenright}\ dir{\isacharparenright}{\isacharparenright}\ {\isasymsqinter}\isanewline
\isaindent{\textsf{else}\ \ \ \ {\isacharbar}\ \ \ \ \ \ {\isacharbar}\ \ \ do\ r\ {\isasymleftarrow}\ }return\ {\isacharparenleft}Error\ eNameTooLong{\isacharparenright}{\isacharsemicolon}\isanewline
\isaindent{\textsf{else}\ \ \ \ {\isacharbar}\ \ \ \ \ \ {\isacharbar}\ \ \ }\ \ \ \textsf{case}\ r\ \textsf{of}\ Error\ e\ {\isasymRightarrow}\ return\ {\isacharparenleft}{\isacharparenleft}afs{\isacharcomma}\ vdir{\isacharcomma}\ vnode{\isacharparenright}{\isacharcomma}\ Error\ e{\isacharparenright}\isanewline
\isaindent{\textsf{else}\ \ \ \ {\isacharbar}\ \ \ \ \ \ {\isacharbar}\ \ \ \ \ \ }{\isacharbar}\ Success\ dir\ {\isasymRightarrow}\isanewline
\isaindent{\textsf{else}\ \ \ \ {\isacharbar}\ \ \ \ \ \ {\isacharbar}\ \ \ \ \ \ {\isacharbar}\ \ \ }do\ r\ {\isasymleftarrow}\ Success\ {\isacharbackquote}\ {\isacharbraceleft}sz\ {\isacharbar}\ v{\isacharunderscore}size\ vdir\ {\isacharless}\ sz{\isacharbraceright}\ {\isasymsqinter}\ return\ {\isacharparenleft}Error\ eOverflow{\isacharparenright}{\isacharsemicolon}\isanewline
\isaindent{\textsf{else}\ \ \ \ {\isacharbar}\ \ \ \ \ \ {\isacharbar}\ \ \ \ \ \ {\isacharbar}\ \ \ }\ \ \ \textsf{case}\ r\ \textsf{of}\ Error\ e\ {\isasymRightarrow}\ return\ {\isacharparenleft}{\isacharparenleft}afs{\isacharcomma}\ vdir{\isacharcomma}\ vnode{\isacharparenright}{\isacharcomma}\ Error\ e{\isacharparenright}\isanewline
\isaindent{\textsf{else}\ \ \ \ {\isacharbar}\ \ \ \ \ \ {\isacharbar}\ \ \ \ \ \ {\isacharbar}\ \ \ \ \ \ }{\isacharbar}\ Success\ newsz\ {\isasymRightarrow}\isanewline
\isaindent{\textsf{else}\ \ \ \ {\isacharbar}\ \ \ \ \ \ {\isacharbar}\ \ \ \ \ \ {\isacharbar}\ \ \ \ \ \ {\isacharbar}\ \ \ }do\ time\ {\isasymleftarrow}\ return\ {\isacharparenleft}v{\isacharunderscore}ctime\ vnode{\isacharparenright}{\isacharsemicolon}\isanewline
\isaindent{\textsf{else}\ \ \ \ {\isacharbar}\ \ \ \ \ \ {\isacharbar}\ \ \ \ \ \ {\isacharbar}\ \ \ \ \ \ {\isacharbar}\ \ \ do\ }dir\ {\isasymleftarrow}\ return\ {\isacharparenleft}dir{\isasymlparr}i{\isacharunderscore}ctime\ {\isacharcolon}{\isacharequal}\ time{\isacharcomma}\ i{\isacharunderscore}mtime\ {\isacharcolon}{\isacharequal}\ time{\isasymrparr}{\isacharparenright}{\isacharsemicolon}\isanewline
\isaindent{\textsf{else}\ \ \ \ {\isacharbar}\ \ \ \ \ \ {\isacharbar}\ \ \ \ \ \ {\isacharbar}\ \ \ \ \ \ {\isacharbar}\ \ \ do\ }inode\ {\isasymleftarrow}\ return\ {\isacharparenleft}afs{\isacharunderscore}inode{\isacharunderscore}from{\isacharunderscore}vnode\ vnode{\isacharparenright}{\isacharsemicolon}\isanewline
\isaindent{\textsf{else}\ \ \ \ {\isacharbar}\ \ \ \ \ \ {\isacharbar}\ \ \ \ \ \ {\isacharbar}\ \ \ \ \ \ {\isacharbar}\ \ \ do\ }{\isacharparenleft}afs{\isacharcomma}\ r{\isacharparenright}\ {\isasymleftarrow}\ afs{\isacharunderscore}update\ afs\ {\isacharparenleft}{\isasymlambda}f{\isachardot}\ f{\isasymlangle}i{\isacharunderscore}ino\ inode\ {\isasymmapsto}\ inode{\isacharcomma}\ i{\isacharunderscore}ino\ dir\ {\isasymmapsto}\ dir{\isasymrangle}{\isacharparenright}{\isacharsemicolon}\isanewline
\isaindent{\textsf{else}\ \ \ \ {\isacharbar}\ \ \ \ \ \ {\isacharbar}\ \ \ \ \ \ {\isacharbar}\ \ \ \ \ \ {\isacharbar}\ \ \ }\ \ \ \textsf{case}\ r\ \textsf{of}\ Error\ e\ {\isasymRightarrow}\ return\ {\isacharparenleft}{\isacharparenleft}afs{\isacharcomma}\ vdir{\isacharcomma}\ vnode{\isacharparenright}{\isacharcomma}\ Error\ e{\isacharparenright}\isanewline
\isaindent{\textsf{else}\ \ \ \ {\isacharbar}\ \ \ \ \ \ {\isacharbar}\ \ \ \ \ \ {\isacharbar}\ \ \ \ \ \ {\isacharbar}\ \ \ \ \ \ }{\isacharbar}\ Success\ {\isacharparenleft}{\isacharparenright}\ {\isasymRightarrow}\isanewline
\isaindent{\textsf{else}\ \ \ \ {\isacharbar}\ \ \ \ \ \ {\isacharbar}\ \ \ \ \ \ {\isacharbar}\ \ \ \ \ \ {\isacharbar}\ \ \ \ \ \ {\isacharbar}\ \ \ }return\ {\isacharparenleft}{\isacharparenleft}afs{\isacharcomma}\ vdir{\isasymlparr}v{\isacharunderscore}ctime\ {\isacharcolon}{\isacharequal}\ time{\isacharcomma}\ v{\isacharunderscore}mtime\ {\isacharcolon}{\isacharequal}\ time{\isacharcomma}\ v{\isacharunderscore}size\ {\isacharcolon}{\isacharequal}\ newsz{\isasymrparr}{\isacharcomma}\ vnode{\isacharparenright}{\isacharcomma}\isanewline
\isaindent{\textsf{else}\ \ \ \ {\isacharbar}\ \ \ \ \ \ {\isacharbar}\ \ \ \ \ \ {\isacharbar}\ \ \ \ \ \ {\isacharbar}\ \ \ \ \ \ {\isacharbar}\ \ \ return\ \ }Success\ {\isacharparenleft}{\isacharparenright}{\isacharparenright}\isanewline
\isaindent{\textsf{else}\ \ \ \ {\isacharbar}\ \ \ \ \ \ {\isacharbar}\ \ \ \ \ \ {\isacharbar}\ \ \ \ \ \ {\isacharbar}\ \ \ }od\isanewline
\isaindent{\textsf{else}\ \ \ \ {\isacharbar}\ \ \ \ \ \ {\isacharbar}\ \ \ \ \ \ {\isacharbar}\ \ \ }od\isanewline
\isaindent{\textsf{else}\ \ \ \ {\isacharbar}\ \ \ \ \ \ {\isacharbar}\ \ \ }od\isanewline
\isaindent{\textsf{else}\ \ \ \ {\isacharbar}\ \ \ }od\isanewline
\isaindent{\textsf{else}\ }od%
\end{isabelle}
\let\isanewline\oldisanewline
\caption{\label{fig:create-corrspec}Functional specification of the \emph{create} operation.}
\end{figure}
The state of the file system, of type~\isa{afs{\isacharunderscore}state}, is passed as the argument~\isa{afs}.
\fsop{create} also takes
the parent directory vnode~\isa{vdir},
the name~\isa{name} of the file to create, a mode attribute~\isa{mode}, and a
vnode~\isa{vnode} to fill with the information of the newly created file. It returns
three values: the updated file system state, the updated parent directory vnode, and
the updated vnode. 

The specification precisely describes the file system behaviour expected by
the VFS, including possible failure modes.
For instance, the implementation needs to return an error if
the file system is in read-only
mode (line 1). On line 2, it allocates a new inode number
and initialises the vnode fields by calling the function \isa{afs{\isacharunderscore}init{\isacharunderscore}inode}
(not shown).
If \isa{afs{\isacharunderscore}init{\isacharunderscore}inode} returns an
\isa{Error} (line 3), \fsop{create} returns \isa{afs}, \isa{vdir}
and \isa{vnode} unchanged as well as the error \isa{Error\ eNFile} indicating
the file system ran out of inode numbers. This pattern is repeated on lines 6, 10, 13
and 19, which each check for errors in preceding operations and specify that these errors
must be propagated to \fsop{create}'s caller, leaving the file system state unchanged.

On line 5, the specification reads the parent directory given by the \isa{vnode} argument,
and computes what the new directory should look like (with the file being created added to it).
Line 12 implies that the size of the directory
must increase. The core of \fsop{create} is specified on line 18 where the file system state
is updated with the updated directory and file inode.%
\end{isamarkuptext}%
\isamarkuptrue%
\isadelimtheory
\endisadelimtheory
\isatagtheory
\endisatagtheory
{\isafoldtheory}%
\isadelimtheory
\endisadelimtheory
\end{isabellebody}%

\vspace{-3ex}
\paragraph{Specifying \fsop{fsync}}
\begin{isabellebody}%
\setisabellecontext{SyncSpec}%
\isadelimtheory
\endisadelimtheory
\isatagtheory
\endisatagtheory
{\isafoldtheory}%
\isadelimtheory
\endisadelimtheory
\isadelimproof
\endisadelimproof
\isatagproof
\endisatagproof
{\isafoldproof}%
\isadelimproof
\endisadelimproof
\isadelimproof
\endisadelimproof
\isatagproof
\endisatagproof
{\isafoldproof}%
\isadelimproof
\endisadelimproof
\isadelimproof
\endisadelimproof
\isatagproof
\endisatagproof
{\isafoldproof}%
\isadelimproof
\endisadelimproof
\isadelimproof
\endisadelimproof
\isatagproof
\endisatagproof
{\isafoldproof}%
\isadelimproof
\endisadelimproof
\isadelimproof
\endisadelimproof
\isatagproof
\endisatagproof
{\isafoldproof}%
\isadelimproof
\endisadelimproof
\begin{isamarkuptext}%
\begin{figure}[tb]
\newcounter{afssync}
\let\oldisanewline\isanewline
\renewcommand\isanewline{\oldisanewline\stepcounter{afssync}\texttt{\arabic{afssync}}\tabto{.75cm}}
\begin{isabelle}%
afs{\isacharunderscore}fsync\ afs\ {\isasymequiv}\isanewline
\textsf{if}\ a{\isacharunderscore}is{\isacharunderscore}readonly\ afs\ \textsf{then}\ return\ {\isacharparenleft}afs{\isacharcomma}\ Error\ eRoFs{\isacharparenright}\isanewline
\textsf{else}\ do\ afs\ {\isasymleftarrow}\ afs{\isacharunderscore}apply{\isacharunderscore}updates{\isacharunderscore}nondet\ afs{\isacharsemicolon}\isanewline
\isaindent{\textsf{else}\ }\ \ \ \textsf{if}\ a{\isacharunderscore}medium{\isacharunderscore}updates\ afs\ {\isacharequal}\ {\isacharbrackleft}{\isacharbrackright}\ \textsf{then}\ return\ {\isacharparenleft}afs{\isacharcomma}\ Success\ {\isacharparenleft}{\isacharparenright}{\isacharparenright}\isanewline
\isaindent{\textsf{else}\ \ \ \ }\textsf{else}\ do\ e\ {\isasymleftarrow}\ select\ {\isacharbraceleft}eIO{\isacharcomma}\ eNoMem{\isacharcomma}\ eNoSpc{\isacharcomma}\ eOverflow{\isacharbraceright}{\isacharsemicolon}\isanewline
\isaindent{\textsf{else}\ \ \ \ \textsf{else}\ }\ \ \ return\ {\isacharparenleft}afs{\isasymlparr}a{\isacharunderscore}is{\isacharunderscore}readonly\ {\isacharcolon}{\isacharequal}\ e\ {\isacharequal}\ eIO{\isasymrparr}{\isacharcomma}\ Error\ e{\isacharparenright}\isanewline
\isaindent{\textsf{else}\ \ \ \ \textsf{else}\ }od\isanewline
\isaindent{\textsf{else}\ }od%
\end{isabelle}
\let\isanewline\oldisanewline
\caption{\label{fig:sync-corrspec}Functional specification of the \emph{fsync} operation.}
\end{figure}

\autoref{fig:sync-corrspec} shows the specification for
\fsop{fsync}, the file system operation that propagates all in-memory updates to the disk.
\fsop{fsync} returns an appropriate error when the file system is in read-only mode. Otherwise,
it propagates the in-memory updates to the medium using the \isa{afs{\isacharunderscore}apply{\isacharunderscore}updates{\isacharunderscore}nondet}
function of \autoref{s:asyncwrites}. Recall that this function applies the first $n$
in-memory updates, with $n$ chosen nondeterministically to model the effect of e.g. a disk
failure happening part-way through. \fsop{fsync} returns successfully when all updates
are applied; otherwise it returns an appropriate error code. When an I/O error occurs
(\isa{eIO}), indicating a storage medium failure, the file system is put into read-only mode to
prevent any further updates to the medium (whose state may now be inconsistent).

The economy of the \fsop{fsync} specification shows the advantage we obtain by carefully
choosing an appropriate representation for the in-memory updates, separate from the on disk
state of the file system.%
\end{isamarkuptext}%
\isamarkuptrue%
\isadelimtheory
\endisadelimtheory
\isatagtheory
\endisatagtheory
{\isafoldtheory}%
\isadelimtheory
\endisadelimtheory
\end{isabellebody}%

\vspace{-3ex}
\section{Limitations\label{s:limitations}}

We conclude with a discussion of the BilbyFs AFS in order to tease out
its limitations and avenues for future improvement. 
An obvious limitation of the AFS is that, like all other file systems specifications of
which we are aware,
it supports no form of concurrency, and so
implicitly specifies that top-level
operations cannot run concurrently to one another. 

Another limitation of the AFS is that it imposes a strict ordering on all updates to
be applied to the storage medium. In practice, many other file systems impose weaker
ordering constraints, especially file systems that are highly concurrent or those built
on top of low-level block interfaces that are asynchronous.
Some file systems can reorder asynchronous writes of data but
not those for meta-data. Investigating how to specify these weaker guarantees
while still retaining the economy and simplicity of the AFS is an obvious
avenue for future work.

The AFS, besides excluding concurrency, also does not specify the interaction
between BilbyFs and the Linux kernel's inode, directory entry and
page caches. The BilbyFs implementation makes use of a small C wrapper,
pictured in \autoref{f:afsoverview},
that sits between it and the Linux VFS, which manages these caches and
implements a global locking discipline that ensures that no two invocations
of the BilbyFs file
system operations can ever run concurrently to each other.
This ensures that the AFS need not concern itself with the aforementioned
Linux caches; however, specifying the behaviour of the Linux caching
layers and their correct interaction with BilbyFs might be another interesting
area for future work.

Another limitation of the AFS arises because function arguments are passed by value
(rather than e.g. as pointers to variables in a mutable heap).
This prevents the AFS from specifying VFS operations that
take multiple pointer arguments that point to the same variable (i.e. are aliases for
  each other). 
Fortunately, the only top-level function able to take arguments that may alias
is the VFS \fsop{rename} operation, which takes two directory pointer arguments that
respectively identify the source
and target directory of the file to be renamed.
When a file is renamed without changing its directory,
the two directory arguments will alias.
We exclude all such aliasing from BilbyFs by having the C wrapper
check for this case: when the two pointers alias it invokes a separate
top-level function of BilbyFs \fsop{rename} that is a special case of the
\fsop{move} operation for when the source and target directory are identical.

A final limitation of the AFS presented here is that, unlike e.g. the recent
work of Chen et al.~\cite{Chen_ZCCKZ_15}, it does not
specify the correct behaviour of the \fsop{mount} operation, which is called at boot time, 
nor specify the file system state following
a crash, for instance to require that the file system is crash-tolerant.

\bibliographystyle{eptcs}
\bibliography{references}
\end{document}